\documentclass[preprint,3p]{elsarticle}

\pdfoutput=1
\usepackage{hyperref}

\journal{JSV}

\biboptions{numbers,sort&compress} 

\usepackage{mathptmx}       

\usepackage{makeidx}         
\usepackage{graphicx}        
\usepackage{natbib}
\usepackage{float}
\usepackage{transparent,color}

\usepackage{amsmath,amsfonts,amssymb}

\usepackage{import}

\usepackage{subcaption}
\usepackage{xspace}

\usepackage{capt-of}
\usepackage{bm}

\usepackage{booktabs}

\usepackage{pgfplots}
\usepgfplotslibrary{external}
\usetikzlibrary{plotmarks,arrows}
\usetikzlibrary{external}
\pgfplotsset{compat=newest}
\usetikzlibrary{calc}

\newlength\fwidth

\definecolor{orange}{rgb}{0.89844,0.37891,0.00391}%
\definecolor{purple}{rgb}{0.36719,0.23438,0.59766}%
\definecolor{gray}{rgb}{0.7,0.7,0.7}%
\definecolor{yellow}{rgb}{0.9922,0.8867,0.5664}%

\definecolor{blue1}{rgb}{0.78040,0.91370,0.70590}%
\definecolor{blue2}{rgb}{0.49800,0.80390,0.73330}%
\definecolor{blue3}{rgb}{0.25490,0.71370,0.76860}%
\definecolor{blue4}{rgb}{0.11370,0.56860,0.75290}%
\definecolor{blue5}{rgb}{0.13330,0.36860,0.65880}%
\definecolor{blue6}{rgb}{0.14510,0.20390,0.58040}%
\definecolor{blue7}{rgb}{0.03140,0.11370,0.34510}%

\newcommand{\bs}{\boldsymbol}

\newcommand{\ie}{i.e.\,}
\newcommand{\eg}{e.g.\,}
\newcommand{\cf}{cf.\,}
\newcommand{\etal}{et\,al.\,}
\newcommand{\sref}[1]{Section \ref{sec:#1}}

\newcommand{\eref}[1]{Eq.\ (\ref{eq:#1})}
\newcommand{\fref}[1]{Fig.\ \ref{fig:#1}}
\newcommand{\tref}[1]{Tab.\ \ref{tab:#1}}
\newcommand{\EPMC}{EPMC\xspace}
\newcommand{\EMA}{LMA\xspace}

\newcommand{\ee}{\mathrm{e}}
\newcommand{\ii}{\mathrm{i}}
\newcommand{\dd}{\mathrm{d}}
\newcommand{\real}[1]{\operatorname{Re}\left\lbrace #1 \right\rbrace}

\newcommand{\abs}[1]{\left| #1 \right|}
\newcommand{\herm}{{}^{\mathrm H}}
\newcommand{\tra}{{}^{\mathrm T}}
\newcommand{\e}[2]{\begin{equation} #1 \label {eq:#2} \end{equation}}
\newcommand{\ea}[2]{
	\begin{eqnarray}
	#1 \label {eq:#2} \end{eqnarray}}

\newcommand{\ForceVec}{\hat{\mathbf f}}
\newcommand{\timevar}{t}
\newcommand{\period}{T}

\newcommand{\mass}{\mathbf M}
\newcommand{\nlforce}{\mathbf g}
\newcommand{\selfexc}{\xi}
\newcommand{\displacement}{\mathbf x}
\newcommand{\velocity}{\dot{\mathbf x}}

\newcommand{\ommod}{\omega}
\newcommand{\Dmod}{\zeta}
\newcommand{\omlin}{\omega_\text{lin}}
\newcommand{\Dlin}{\zeta_\text{lin}}
\newcommand{\shpmod}{\hat{\mathbf x}}

\newcommand{\shpmodlinmatrix}{\bm\Phi_\text{lin}}
\newcommand{\shpmodnorm}{\bm\phi}
\newcommand{\modamp}{a}
\newcommand{\phasemod}{\theta_m}
\newcommand{\phaselag}{\theta}

\makeindex             

\begin{document}

\begin{frontmatter}
\title{Challenging an Experimental Nonlinear Modal Analysis Method with a New Strongly Friction-damped Structure}

\author[addressILA]{Maren Scheel}
\author[addressTobi]{Tobias Weigele}
\author[addressILA]{Malte Krack}

\address[addressILA]{University of Stuttgart, Pfaffenwaldring 6, 70569 Stuttgart, Germany; scheel@ila.uni-stuttgart.de; krack@ila.uni-stuttgart.de} 
\address[addressTobi]{MesH Engineering GmbH, Schwertstra\ss e 58-60, 71065 Sindelfingen, Germany}

\begin{abstract}
In this work, we show that a recently proposed method for experimental nonlinear modal analysis based on the extended periodic motion concept is well suited to extract modal properties for strongly nonlinear systems (\ie in the presence of large frequency shifts, high and nonlinear damping, changes of the mode shape, and higher harmonics). To this end, we design a new test rig that exhibits a large extent of friction-induced damping (modal damping ratio up to 15 \%) and frequency shift by 36 \%. 
The specimen, called RubBeR, is a cantilevered beam under the influence of dry friction, ranging from full stick to mainly sliding. With the specimen's design, the measurements are well repeatable for a system subjected to dry frictional force. 
Then, we apply the method to the specimen and show that single-point excitation is sufficient to track the modal properties even though the deflection shape changes with amplitude.
Computed frequency responses using a single nonlinear-modal oscillator with the identified modal properties agree well with measured reference curves of different excitation levels, indicating the modal properties' significance and accuracy.
\end{abstract}

\begin{keyword}
experimental modal analysis, nonlinear modes, jointed structures, friction damping, force appropriation
\end{keyword}

\end{frontmatter}

\section{Introduction}

During operation, machines undergo dynamic loading causing vibrations. 
To ensure safe operation and a long product life, vibration amplitudes must be limited, which necessitates damping. 
Material damping of light-weight structures, for example in turbo-machinery applications, is often negligible. Thus, dry friction damping is introduced as the main means for vibration reduction \cite{Griffin1990}, \eg in the form of under-platform dampers for turbine blades. Dry friction is, as almost all damping mechanisms, inherently nonlinear (\ie the force-deflection relationship is nonlinear) \cite{Popp2003}. 
To accurately predict vibrations under the influence of dry friction, computational and experimental methods are required that properly account for nonlinearities. Based on experimental data, computational models are validated or updated. Furthermore, experimental methods serve as means for structural health monitoring or controller design. In the last years, various approaches have been suggested and are currently explored and further developed \cite{Kerschen2006,Noel2017b}.
In this work, we assess the limitations and capabilities of a nonlinear experimental method that extends the idea of modal testing (\ie experimentally determining the modal properties frequencies, damping ratios, and deflection shapes) to structures and mechanical systems, subjected to nonlinear forces.

Commonly, modal testing is conducted assuming linear vibration behavior. This approach is subsequently abbreviated as \EMA for linear modal analysis. Further, the mode of a linear system is denoted linear mode, and its properties are denoted accordingly (\eg linear mode shape). 
In the presence of nonlinear forces, conducting \EMA can lead to the identification of spurious modes, wrong damping measures or misinterpreting nonlinear effects as noise or variance in the measurements \cite{Ewins2000}.
To properly account for nonlinearities, an extension of the concept of modes to nonlinear systems was proposed by defining nonlinear modes with amplitude-dependent modal properties.
According to the most common definition \cite{Rosenberg1960}, the nonlinear mode is a family of periodic motions of a conservative, autonomous system. This concept no longer applies to systems with nonlinear damping, for example friction-damped structures. 
Therefore, the notion of a nonlinear mode as periodic motion has been extended \cite{Krack2015}, defining a nonlinear mode as a family of periodic motions of the \textit{damped} system. Periodicity is enforced with an artificial negative damping term that compensates the energy loss due to natural dissipation over one period.

To identify amplitude-dependent modal properties from experimental data (\ie nonlinear modal testing), \EMA methods must be extended or new approaches must be found. One strategy is to identify a model replicating the measured nonlinear dynamics, using a nonlinear system identification approach. Based on this model, modal properties are obtained in a second step, using a numerical nonlinear modal analysis method. This was, for example, suggested using frequency-based nonlinear system identification \cite{Noel2016b}. This method, like many nonlinear system identification methods, requires an a-priori choice of nonlinear basis functions. Choosing suitable basis functions can be challenging, \eg for dry friction. For such nonlinearities, frequency-based nonlinear system identification is still in its infancy \cite{Noel2016b}.
The identified models depend on the choice of basis functions and model order \cite{Noel2016b}. Due to inevitable modeling errors, the identified models also depend on the used training data. Thus, great care must be taken to identify an accurate model and obtain meaningful modal properties.

Amplitude-dependent damping ratios of bolted joints are often identified by measuring a hysteresis loop and determining the associated dissipated work in the joint \cite{Bograd2011,Scheel2018}. Alternatively, hysteresis measurements can be used to update joint model parameters \cite{Willner2018}. Hysteresis-based approaches are, however, only applicable to local nonlinearities since the local nonlinear forces and relative displacement at the nonlinearity must be measured. 

A different approach for nonlinear modal testing, suited for both conservative and non-conservative nonlinearities is adopted from modal hammer testing. By analyzing the free decay, amplitude-dependent modal properties are extracted for the vibration levels covered by the free decay \cite{Kuether2016,Sracic2012,Stephan2017}. This approach assumes weak nonlinearity such that modes remain uncoupled, even when several modes are simultaneously excited through impulse forcing. For stronger nonlinearities, however, inevitable mode coupling and the lack of superposition hamper the correct extraction of modal properties of a single mode \cite{Kuether2016}.
To overcome this difficulty, force appropriation can be used to excite only a single mode. After removing the excitation, the free decay is analyzed to obtain the modal properties \cite{Peeters2011a,Heller2009,Londono2015}. Removing the excitation, however, often distorts the measurement as the removal event usually causes a sudden force pulse or the passive shaker adds mass \cite{Dion2013a}. 

For structures subjected to high damping, the oscillation decays quite quickly. In such cases, noise at low amplitudes and the finite time-frequency resolution limit the accuracy of free-decay approaches.
An alternative is to use force appropriation to excite periodic motion, which allows for a straight-forward signal analysis with a fine frequency and amplitude resolution.
This approach is a consistent extension of linear phase resonance testing.
Force appropriation is commonly ensured using controlled excitation, for example with control-based continuation \cite{Renson2016b,Renson2017} or a phase-locked loop (PLL). The latter approach was successfully applied to numerous structures with both conservative and nonconservative nonlinearities \cite{Denis2018,Peter2017,Scheel2018,Scheel2020,Schwarz2019}, such as a Chinese gong, a beam with magnetic nonlinearity, the joint resonator, or the Brake-Reu\ss-beam. These test specimens were excited ensuring phase resonance between the fundamental harmonic content of excitation and response.
Frequency shifts of up to 25 \% were tracked \cite{Denis2018}, but damping was small with damping ratios up to 1.1 \% \cite{Scheel2018}. No significant change in the deflection shape has been reported for these systems.

With free-decay approaches, amplitude-dependent modal damping ratios of up to 8 \% were determined experimentally, where damping was caused by dry friction \cite{Kuether2016,Heller2009,Stephan2017,Londono2015}. Virtual experiments suggest that high damping ratios can be extracted with force appropriation using PLL \cite{Scheel2018}, but this has not yet been experimentally verified. 
Therefore, the aim of this work is to experimentally test the opportunities and limitations of force appropriation using PLL in case of strong damping and a large frequency shift. We assess if shaker-stinger excitation at only one location with a single controlled harmonic is sufficient to track modal properties for a strong damping nonlinearity.
To asses the accuracy of the identified modal properties, we predict frequency responses around the resonance and compare them to measured reference curves. For the prediction of frequency response, we assume that the dynamics are captured by a single nonlinear mode with the identified modal properties.

For this study, a test specimen is needed, which features strong changes in resonance frequency and high damping. Nonlinear forces due to dry friction cause significant changes in both resonance frequency and modal damping ratio, when the contact dynamics change from full-stick to mainly sliding.
Several test rigs with friction-induced damping have been presented in the past.
A three-parts beam with planar joints was suggested \cite{Peyret2010,Dion2013a} to study dry friction in a general setting. This specimen, however, is designed for the measurement of micro-slip effects and does not allow for macro-slip. 
Alternative test rigs replicate a specific joint or contact situation, for example under-platform damper test rigs (\eg \cite{Sanliturk2001,Berruti2011,Pesaresi2016}) or bolted joint such as the Brake-Reu\ss-beam \cite{Brake2019}, the joint resonator \cite{bohl1987,Segalman2009} or the H testbench \cite{Stephan2017}.
Since the authors are not aware of a suitable test rig that covers the full range from full-stick to mainly sliding, we suggest a new design of a cantilevered beam under the influence of dry friction. The design is modular such that the location of the friction joint can be varied, allowing for varying severeness of the nonlinearity. Additionally, the design focuses on achieving well repeatable measurements.

The paper is structured as follows. In \sref{method}, the \EPMC and the nonlinear modal testing approach is summarized. Then, the specimen's design is explained in \sref{design}, followed by the results of the linear and nonlinear experimental modal analysis in \sref{results}. The paper closes with conclusions in \sref{conclusion}.

\section{Nonlinear Modal Testing Method}\label{sec:method}

First, we briefly summarize the theory of the extended periodic motion concept (\EPMC) \cite{Krack2015}. Then, a simple experimental implementation to isolate a nonlinear mode and identify its properties, introduced in \cite{Scheel2018}, is recalled.

\subsection{The Extended Periodic Motion Concept (\EPMC)}\label{sec:epmc}

We limit the discussion to nonlinear modes that depart from a specific linear mode with a distinct, nonzero modal frequency. In such cases, a nonlinear mode is a continuous extension of a linear mode with respect to an amplitude measure (\eg energy). A nonlinear mode according to the \EPMC is defined as a family of periodic motions of an autonomous, non-conservative system \cite{Krack2015}. To obtain periodic motions, the energy lost by natural dissipation is compensated by an artificial, negative, mass-proportional viscous damping term. 

We consider a system described by a set of $N$ generalized coordinates $\displacement \in \mathbb{R}^N$. The system's dynamics are governed by the equation of motion
\e{\mass \ddot{\displacement} + \nlforce (\displacement,\velocity) = \bs 0.}{ode}
Herein, $\mass$ is the symmetric and positive definite mass matrix, and $\nlforce$ comprises linear and nonlinear restoring and damping forces. The derivative with respect to time $\timevar$ is indicated with the over-dot. To obtain a periodic motion $\displacement(\timevar) = \displacement (\timevar + \period)$ with fundamental period $\period$, the initial system \eref{ode} is augmented with an artificial, negative viscous damping term,
\e{\mass \ddot{\displacement} + \nlforce (\displacement,\velocity) - \selfexc \mass \velocity = \bs 0.}{nm}
The fundamental angular frequency $\ommod = 2\pi/\period$ is defined as the modal frequency. The Fourier coefficients $\shpmod_n \in \mathbb{C}^N$ of $\displacement(\timevar)$ are defined as the Fourier coefficients of the modal deflection shape. The modal damping ratio $\Dmod$ is related to $\selfexc$ via $\selfexc = 2 \Dmod \ommod$. These modal properties (frequency, deflection shape, and damping ratio) depend on the vibration amplitude.

The \EPMC is consistent with the periodic motion concept for conservative systems. Further, it is consistent to the linear case under modal damping, since linear mode shapes are orthogonal with respect to the mass matrix and thus also to the artificial damping term. If two or more linear mode shapes contribute strongly in the nonlinear mode and, at the same time, damping is not light (\ie $|\Dmod|$ not $ \ll 1$), the artificial term will generally distort the modal coupling. Then, the relevance of the nonlinear modal properties to the initial system \eref{ode} is not clear.

When the initial system \eref{ode} is externally excited near a primary resonance, the vibration energy is confined to the corresponding nonlinear mode. Then, the system behaves like a single-degree-of-freedom oscillator, and the resonant frequency response closely follows the frequency-amplitude relation of the nonlinear mode. This holds analogously in case of self-excitation under negative damping of a particular mode \cite{Krack2015}.

\subsection{Experimental Implementation Using Force Appropriation}\label{sec:forcing}

In an experiment, the artificial term in the \EPMC could be replaced by velocity-proportional external forcing. It is, however, practically infeasible to apply forces at all material points of a structure. Therefore, external forcing must be found that replicates the artificial viscous damping term closely, while being feasible to experimentally implement.
For structures with diagonal-dominant mass matrix (\ie weak inertia coupling), such as slender structures, or in case of phase-synchronous motion (for conservative or lightly damped structures), the excitation can be simplified.
Then, an excitation force applied at only one location is already sufficient to isolate the mode \cite{Scheel2018}. Furthermore, it is sufficient for many structures to ensure local phase resonance with respect to only the fundamental harmonic of force and response at the excitation location (drive-point response).

In this work, local phase resonance is ensured using a PLL controller (see \eg \cite{Mojrzisch2016,Peter2017,Denis2018}).
The core element of the PLL is the phase detector (see \fref{pll_scheme}) that determines the phase lag between excitation and response signal. Several implementations can be found in literature \cite{Twiefel2008}. In this work, synchronous detection \cite{ZurichInstruments2016} is used due to its good accuracy in case of multi-harmonic signals (in contrast to the mixing phase detector used in \cite{Scheel2018}).
Using synchronous detection, the phase lag $\hat{\phaselag}_a$ between the fundamental harmonic of measured response and force is determined. A standard PI controller then minimizes the error $e(\timevar) = \hat{\phaselag}_a(\timevar) - \phaselag_a$, \ie the difference between current phase lag and desired phase lag $\phaselag_a$ (set point). To this end, the controller adapts the excitation frequency $\Omega(\timevar)$ until the desired phase lag is reached (\eg $\phaselag_a = \pi/2$ between drive-point acceleration and force for phase resonance). The frequency change $y$ (output of the PI controller) is added to the constant center frequency $\omega_m$ and then integrated, which yields the reference phase $\alpha_R$,
\begin{align}
	y(\timevar) &= K_p e(\timevar) + K_i \int_{0}^{t} e(\tau) \dd \tau\\
	\alpha_R(\timevar) &= \int_{0}^{t} \Omega(\tau) \mathrm{d} \tau = \int_{0}^{t} (y(\tau) + \omega_m) \dd \tau 
\end{align}
This phase is the argument of a cosine function, used as excitation signal with a given amplitude. 
After transients have decayed, the periodic motion of the structure is recorded for this excitation amplitude. Then, the amplitude is stepwise increased or decreased. This way, the phase-resonant points are tracked for different amplitudes (\ie the phase-resonant backbone) \cite{Denis2018,Peter2017,Scheel2018}. The frequency of each periodic motion corresponds to the modal frequency at the associated response amplitude \cite{Krack2015}. 

\begin{figure}
	\centering
	\includegraphics{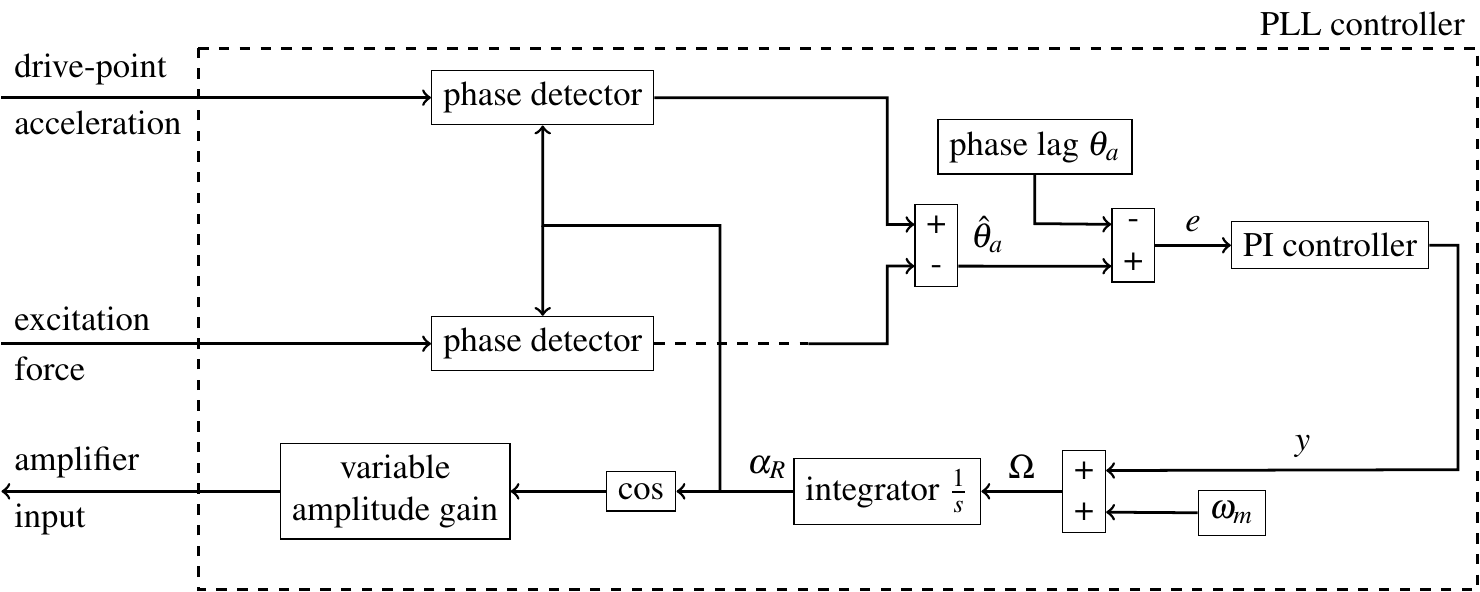}
	\caption{Scheme of the PLL controller with synchronous detection.}
	\label{fig:pll_scheme}
\end{figure}

\subsection{Identification of Modal Properties}

The modal frequency $\ommod$ as well as the Fourier coefficients of the deflection shape $\lbrace \shpmod_0\, , \shpmod_1\, , \shpmod_2\, , \shpmod_3\, , ... \rbrace$ are extracted directly from the measured time signals.
The modal damping ratio is obtained by the balance of excitation power and dissipated power (which has to hold over one cycle of vibration) \cite{Scheel2018}.
With the active excitation power $P_1$ associated with the first Fourier coefficient of excitation force and drive-point velocity, the modal damping ratio is
\e{\Dmod = \dfrac{P_1}{\ommod^3 \modamp^2}.}{dmod}
Here, $\modamp \in \mathbb{R}$ is the magnitude of the modal amplitude, which is the scaling factor between the first Fourier coefficient of the mass-normalized deflection shape, $\shpmodnorm_1$, and the unscaled shape, $\shpmod_1 = \modamp \shpmodnorm_1$ \cite{Scheel2018}.
To compute $\modamp$, linear mass-normalized mode shapes $\shpmodlinmatrix$ are used and obtained with \EMA at low excitation level to estimate the mass matrix with $\left( \shpmodlinmatrix \tra \right)^+ \left(\shpmodlinmatrix \right)^+$. Then, $\modamp$ is determined as \cite{Scheel2018}
\e{\modamp^2 = \shpmod_1\herm \left( \shpmodlinmatrix \tra \right)^+ \left(\shpmodlinmatrix \right)^+ \shpmod_1.}{modalampl}
Here, ${(\bullet)}\tra$ is the transpose, ${(\bullet)}^+$ is the pseudoinverse, and ${(\bullet)}\herm$ is the Hermitian transpose.

\section{The Design of RubBeR}\label{sec:design}

The newly designed specimen is a cantilevered beam called Rubbing Beam Resonator (RubBeR). To serve as suitable test structure for assessing the capabilities and limitation of the nonlinear modal testing method, the following features are required:

\begin{itemize}
	\item The specimen has well isolated modes.
	\item It is possible to obtain well repeatable measurements in order to acquire a solid basis for assessing the experimental method. Especially, the contact situation (\ie contact area and normal preload) is well defined.
	\item It is possible to vary the location of the joint in order to study different configurations with varying significance of the nonlinearity.
	\item It is possible to vary the normal preload at the contact in order to change the amplitude regime where mainly sliding occurs. This way, the setup can be tuned to make full use of the available shaker's force range. 
	\item The structural integrity (\eg with respect to fatigue) during testing is ensured.
	\item Additional joints are avoided that could introduce further nonlinearity.
	\item It is possible to vary the contact interface and partners, both in shape and material. Furthermore, contact partners can be replaced in case of wear.
	\item The specimen's mode shapes exhibit simple dynamics that can be measured with standard equipment (\eg bending modes in only one direction).
\end{itemize}

\begin{figure}
	\centering
	\begin{subfigure}[b]{0.49\textwidth}
		\def\svgwidth{0.9\textwidth}
		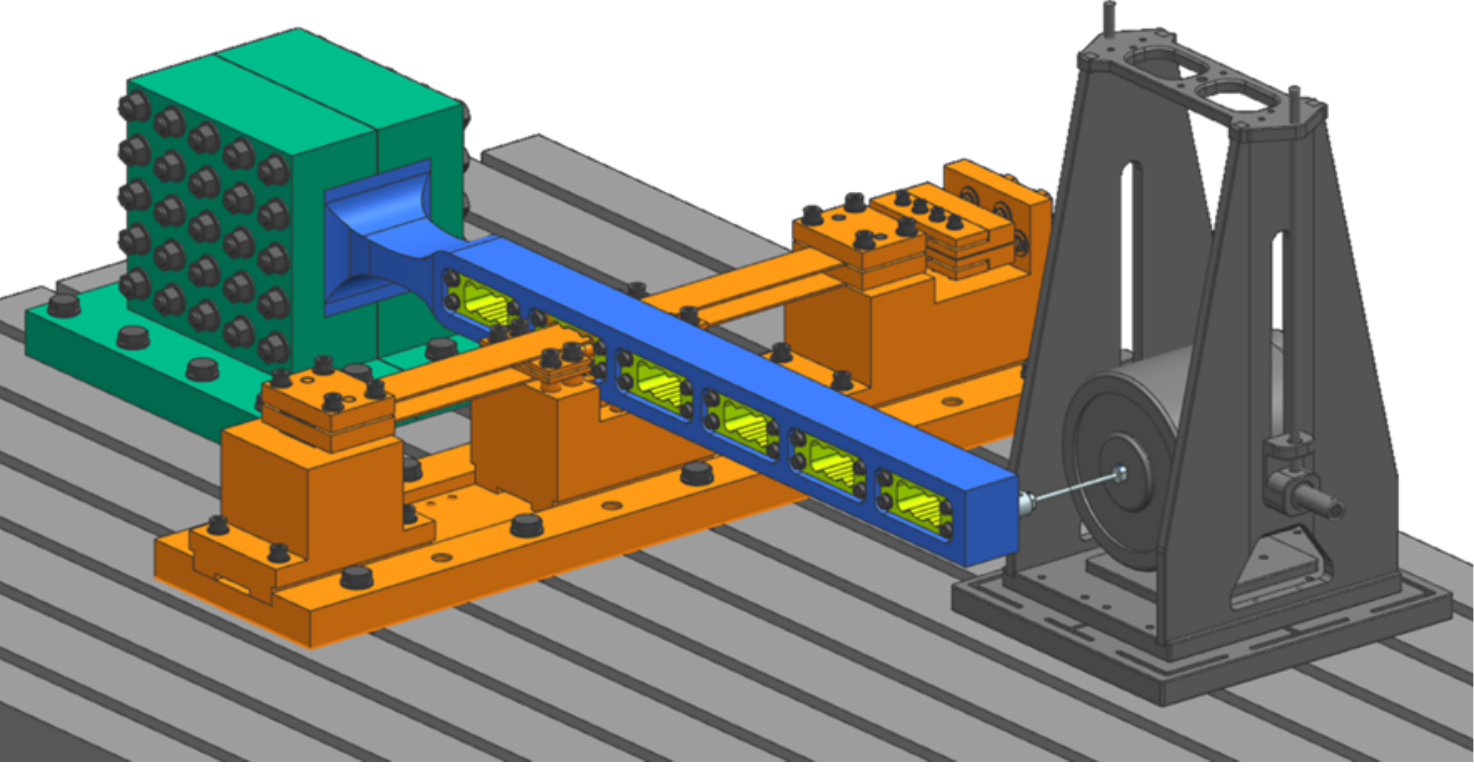
		\caption{}
		\label{fig:setup_cad}
	\end{subfigure}
	\begin{subfigure}[b]{0.49\textwidth}
		\def\svgwidth{0.9\textwidth}
		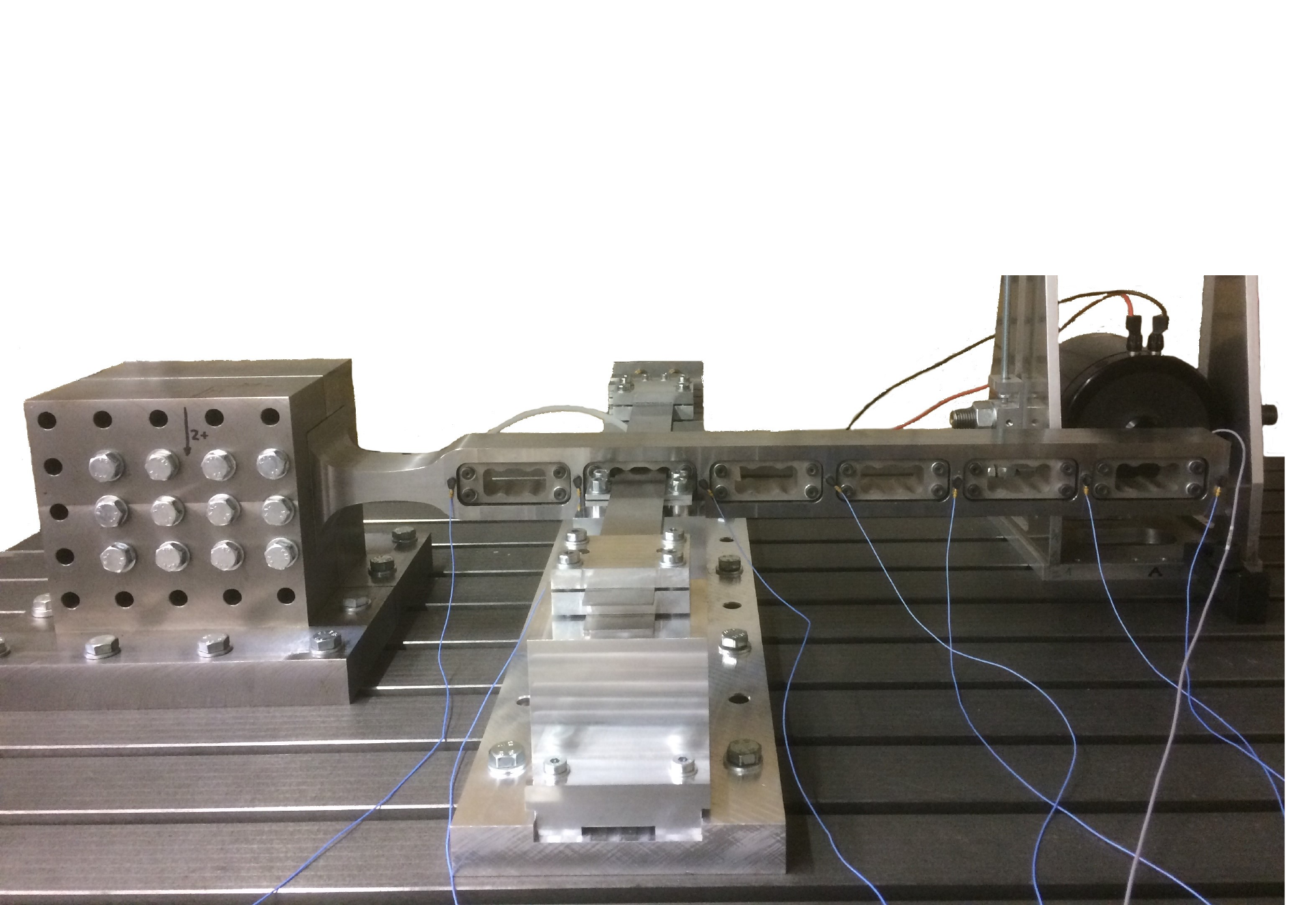
		\caption{}
		\label{fig:setup_experimental}
	\end{subfigure}
	\caption{(a) CAD model and (b) photo of the experimental setup including shaker and sensors.}\label{fig:setup}
\end{figure}

The developed specimen is depicted in \fref{setup_cad}.
The beam (blue) is clamped on one side and is designed to study horizontal bending motion. Its free length is $l = 710 \text{ mm}$, and the cross section at the tip is 60 mm by 50 mm (height by width). 
To avoid exciting the vertical bending mode due to imperfect excitation (\eg a slight misalignment of the stinger), the frequencies of the first bending modes in both directions must be well separated. A lower vertical bending frequency is desired, such that the vertical mode cannot be excited by higher harmonics of the horizontal bending mode.
Therefore, the characteristic fillet close to the clamping was introduced. (The achieved linear modal frequencies of the first bending modes are reported in \sref{linear}.)

The clamping is applied through two outer parts (dark green). 
To prevent stress concentration at the outer edges of the clamping block, the beam is rounded just before the clamping block. The beam's root, \ie the part of the beam that is clamped, is broadened such that twelve drill holes go through the beam (see \fref{beam_clamping}).
By separating the beam from its clamping (instead of manufacturing it as one part), it is easier and more cost efficient to replace the beam in case of damage or in case a different beam geometry is of interest. The beam and clamping are made of milled steel 16MnCr5, which has a high bending fatigue strength compared to carbon steel.
There is contact between beam and clamping block at the two surfaces with the drill holes (gray-shaded area in \fref{beam_clamping}) and additionally, at three lines orthogonal to the two surfaces (red circles in \fref{beam_clamping}).
The contact pressure between beam and clamping is applied through screws whose number can be varied. Preliminary simulations revealed that using the twelve mid screws (see \fref{setup_experimental}) leads to a nearly uniform pressure distribution.
In case of relative motion between clamping and beam at the clamping interface, this motion is orthogonal to the horizontal motion of the beam. Therefore, the induced friction does not influence the friction effects of interest. In case of relative motion between clamping and beam at the additional three contact lines, the influence of friction is expected to be small due to the small contact area.

\begin{figure}
	\centering
	\def\svgwidth{0.55\textwidth}
	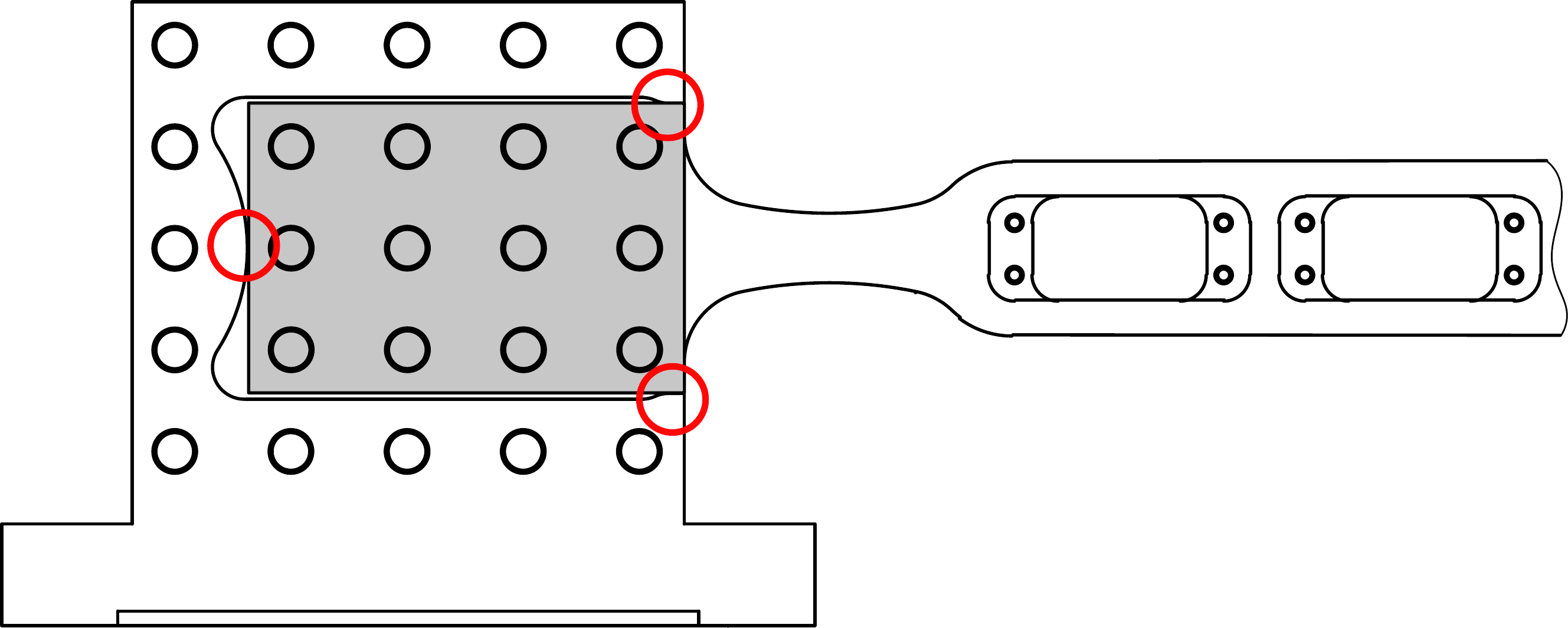
	\caption{Beam and one part of the clamping block. The main contact area between beam and clamping block is the gray-shaded area on both sides of the beam. Additionally, the beam is in contact with the clamping block at three lines, indicated by red circles (oriented orthogonally to the depicted plane).}\label{fig:beam_clamping}
\end{figure}

The beam has six "pockets", distributed over its length. These are potential locations for the joint causing dry friction nonlinearity.
In each of the pockets, inserts (light green) can be attached that serve as the contact partner on the beam side. The inserts are made of the same steel as the beam and are manufactured with wire erosion. This process yields a better surface finish compared to machining. Having inserts (instead of machining the contact area directly to the beam) allows for a modular setup, \ie the inserts can be replaced as required.
The inserts are screwed to the beam; only the surfaces with the screw holes have contact with the beam. Analogously to the clamping of the beam, the contact surface is orthogonal to the horizontal bending motion to not affect the friction effects of interest.

Dry friction is introduced through contact between the inserts and a fixed contact partner.
To achieve repeatable measurements, the contact must be well-defined with respect to the location and area. This can be ensured by a nominal line contact. Therefore, the inserts have four rounded bosses (see \fref{friction_nonlinearity}), whose ridges define the contact area. Two plates of steel 90MnCrV8 (orange in \fref{setup_cad}) are inserted in one of the pockets and touch the upper and lower bosses (see \fref{friction_nonlinearity}). 
The plates are fixed at both ends with the orange clamping device (see \fref{setup_cad}).
In case the plates exhibit undesired dynamics in the frequency range of interest, the length of the metal plates and thus their modal frequencies can be varied. Moreover, the plates are axially pre-stressed to prevent buckling. 

\begin{figure}
	\centering
	\begin{subfigure}[b]{0.45\textwidth}
		\includegraphics[width=0.9\textwidth]{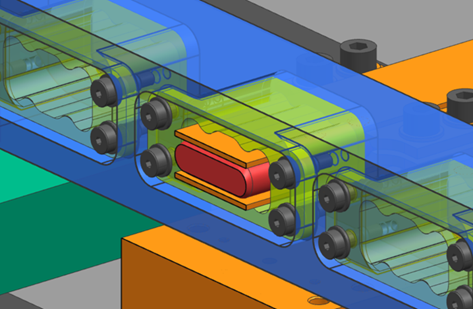}	
		\caption{}\label{fig:friction_nonlinearity}
	\end{subfigure}
	\begin{subfigure}[b]{0.45\textwidth}
		\includegraphics[width=0.9\textwidth]{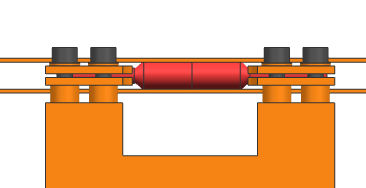}	
		\caption{}\label{fig:airpillow}
	\end{subfigure}
	\caption{(a) CAD model detail of the friction interface with steel plates (orange), air pillow (red), inserts (green) and beam (blue). (b) CAD model detail of the steel plates and air pillow.}\label{fig:cad_detail}
\end{figure}

If the beam vibrates in horizontal direction, relative motion between inserts and plates will lead to dry friction. With the chosen material combination, the contact partner that is cheaper to manufacture (\ie the metal plates) is more prone to wear.
In case of imperfections in the contact zone, it is possible that a sliding motion is initiated first at the upper or lower contact area. This would cause a torsional motion of the beam. Due to the pocket design, the lever arm measured from the beam's midspan is minimized and thus also potential torsional motion. 

To apply the normal load at the contact, an air-filled rubber pillow (red in \fref{cad_detail}) is inserted between the two steel plates. The pillow is a piece of bicycle tube, glued at both ends and clamped additionally.
The pillow ensures that the normal load is homogeneously distributed over the contact area (in the nominal case, not considering machining tolerances). The normal load can be set by changing the air pressure. The relation between air pressure and normal load has been calibrated with a ME-Me\ss systeme K3D60a force sensor. Pressures between 0.033 bar and 0.364 bar (3300 Pa and 36400 Pa) have been calibrated, corresponding to total contact forces at the interface between 10 N and 98 N.
The air pressure-preload relation is nearly linear.
The clamping of the metal plates and the air pillow (see \fref{airpillow}) is designed to constrain horizontal motion, so that the air pillow does neither contribute damping nor stiffness in the direction of interest.

The two linear limit cases of a dry friction contact under constant normal preload are full stick for small vibration levels and full slip for large vibration levels. The deflection shape changes with the vibration level. For small vibration levels, it resembles the mode shape with fixed boundary condition at the friction contact. For the limit case at large vibration levels, it resembles the mode shape with free boundary condition at the friction contact.
Accordingly, the modal frequency decreases with increasing vibration level. In the RubBeR setup, the severeness of the frequency and deflection shape change depends on the location of the steel plates and is more substantial for contact interfaces close to the beam's free end.

\section{Modal Analysis of the RubBeR Test Rig}\label{sec:results}

The setup's instrumentation is schematically shown in \fref{instrumentation}. Seven PCB 352C22 accelerometers are attached to the beam with wax. Note that the accelerometers and the force cell are attached to opposite sides of the beam. Thus, the measured acceleration is not the drive-point acceleration, but is a close approximation.
All six inserts are attached to the pockets. In this work, only the configuration with contact at pocket two is studied. 
In the following, the experiments for the linear and nonlinear modal analysis and the obtained results are discussed.

\begin{figure}
	\centering
	\def\svgwidth{0.9\textwidth}
	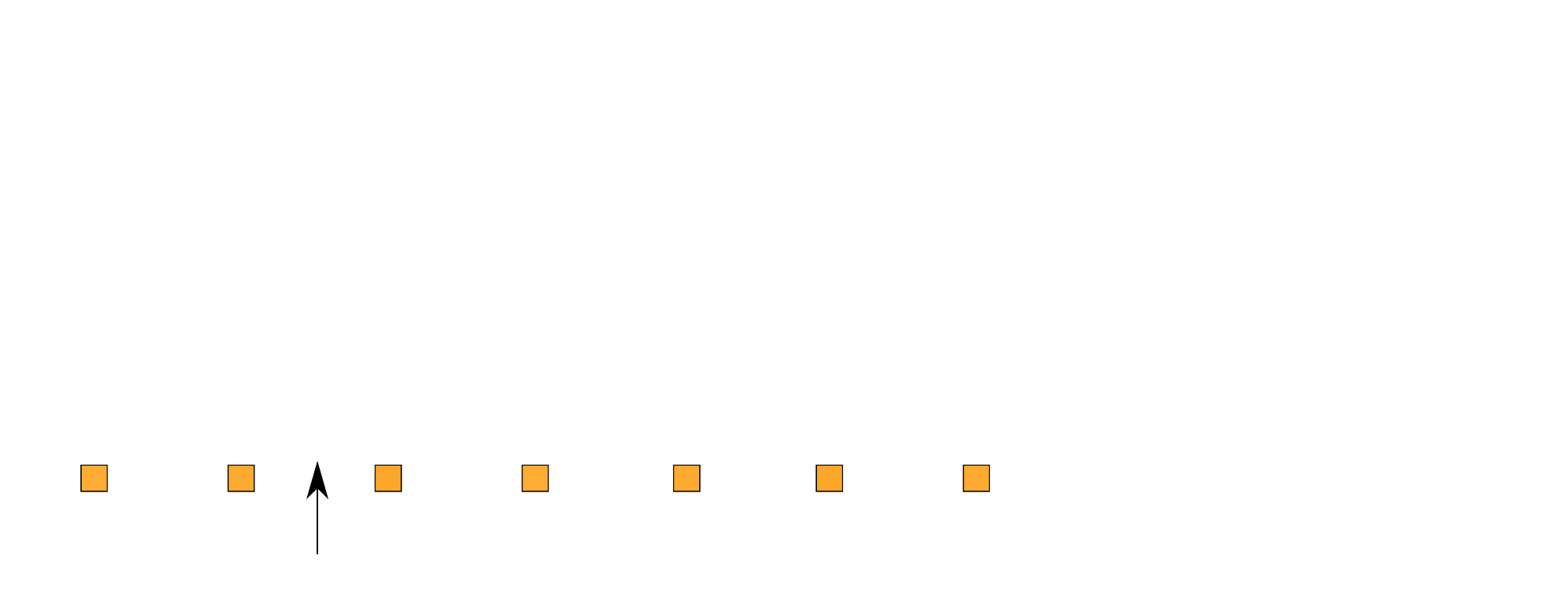
	\caption{Scheme of the instrumented test rig (top view). Two excitation locations are considered in this study, one at the tip (marked with purple) and one at approximately two-third of the beam's length (marked with orange).}
	\label{fig:instrumentation}
\end{figure}

\subsection{Linear Modal Analysis}\label{sec:linear}

First, the configuration without steel plates is studied. First, the beam is excited at the tip in vertical direction using a modal hammer (PCB 086C01). The shaker depicted in \fref{instrumentation} was detached in all hammer tests. For this measurement, an additional accelerometer is attached to the top side close to the tip. The first two vertical bending modes are at 51.9 Hz and 462.32 Hz, respectively. All linear analyses were conducted using m + p Analyzer with the hardware m + p VibRunner.
Next, horizontal bending modes are excited with hammer impacts applied on the beam's midspan close to the location of the force transducer in \fref{setup_cad}. The first bending mode is identified at 67.4 Hz. 
Thus, the beam's design separates the natural frequencies of the first two vertical and the first horizontal bending modes as intended (\cf \sref{design}).

Then, the metal plates are inserted and the air pillow is filled with 0.39 bar. A Br\"{u}el \& Kj\ae r vibration exciter (Type 4809)  is attached to the beam's tip via a steel stinger of 3 mm diameter and 60 mm length and a Br\"{u}el \& Kj\ae r 8230 force transducer (see \fref{setup} and \fref{instrumentation}). A \EMA is conducted for the horizontal bending modes with low-level random excitation from 10 Hz to 3200 Hz. The frequencies and modal damping ratios of the first three bending modes are given in \tref{linear_test}. 
The resonance peak of the first mode is slightly distorted in the low-level frequency response functions, which could indicate nonlinear effects due to small relative motion at the joint. Other sources for damping are the clamping of the beam as well as the suspension of the measurement table.

\begin{table} 
	\centering
	\begin{tabular}{l l l}
		\toprule
		& $\omlin$ in Hz & $\Dlin$ in \% \\
		\midrule[0.8pt] 
		first horizontal bending mode & 111.3 & 0.12 \\
		second horizontal bending mode & 582.3 & 0.41 \\
		third horizontal bending mode & 1096.1 & 0.46 \\
		\bottomrule
	\end{tabular}
	\caption{Modal frequencies and modal damping ratios of the first three horizontal bending modes with contact at the joint, identified with random noise excitation at low level.} \label{tab:linear_test}
\end{table}

\subsection{Nonlinear Modal Analysis}\label{sec:NMA}

The amplitude-dependent modal properties of the first horizontal bending mode are determined with the approach described in \sref{method}. The controller is implemented on a dSPACE MicroLabBox. The control parameters were $\omega_m = 230 \pi\; \text{rad}\: \mathrm{s}^{-1}$, the gains of the proportional and integral part of the PI controller were $K_p =15 \; \mathrm{s}^{-1} $, and $K_i = 3\pi \; \mathrm{s}^{-2}$ (\cf \fref{pll_scheme}). Synchronous detection was implemented using a first-order low-pass filter with the time constant $2/\pi \; \mathrm{s}$. The sampling frequency was 10 000 Hz, which is high enough to sample one period with at least 90 points.
Nonlinear modal testing was performed for two different excitation locations: at the tip (marked with purple in \fref{instrumentation}) and at about two third of the beam's length (marked with orange). This color-coding will be followed for the remainder of the paper. 

The extracted modal frequencies and modal damping ratios for both excitation locations are shown in \fref{modal_properties_freqdamp}, plotted against the normalized fundamental content of the tip deflection. These normalized deflection amplitudes correspond to measured acceleration amplitudes ranging from $0.5 \; \mathrm{m}\: \mathrm{s}^{-2}$ to $40 \; \mathrm{m}\: \mathrm{s}^{-2}$. The modal properties were obtained for amplitudes of the first Fourier coefficient of the force from 0.01 N to 37.16 N and from 0.004 N to 22.3 N for excitation at two-third of the beam's length and at the tip, respectively.
In the measured amplitude range, the well-known amplitude-dependence of the modal properties are observed: The frequency decreases with increasing amplitude, and the modal damping ratio first increases and subsequently decreases for large amplitudes. Here, the frequency decreases for normalized amplitudes above $2\cdot 10^{-5}$ by about 36~\%. The modal damping ratio is about 0.1~\%  for low amplitudes, reaches the maximum value of 15~\%, and then decreases to 7.9~\%. To the best of the authors' knowledge, experiments exhibiting such a large increase in the damping ratio with subsequent decrease are scarce. Stephan \etal reported an increase with subsequent decrease in the modal damping ratio for large vibration amplitudes (from about 5.6~\% to about 4~\%) \cite{Stephan2017}.

\begin{figure}
	\centering
	\begin{subfigure}[b]{0.49\textwidth}
		\centering
		\includegraphics{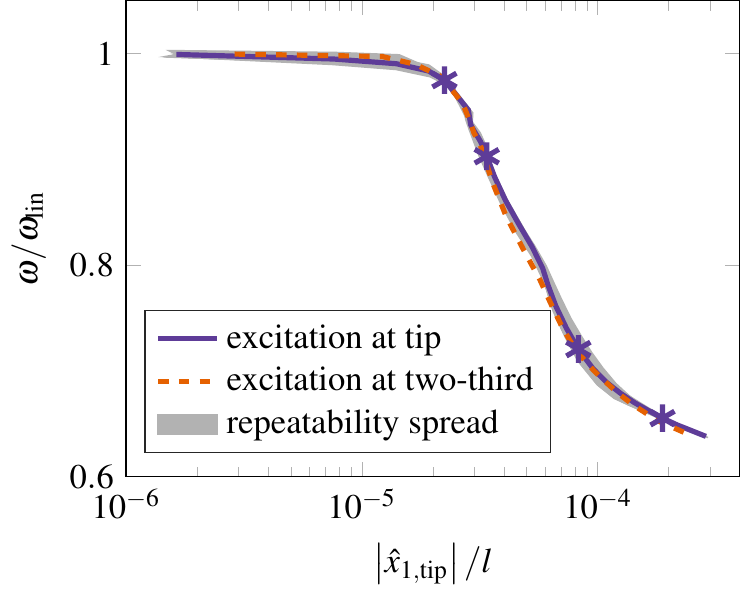}
		\caption{}
	\end{subfigure}
	\begin{subfigure}[b]{0.49\textwidth}
		\centering
		\includegraphics{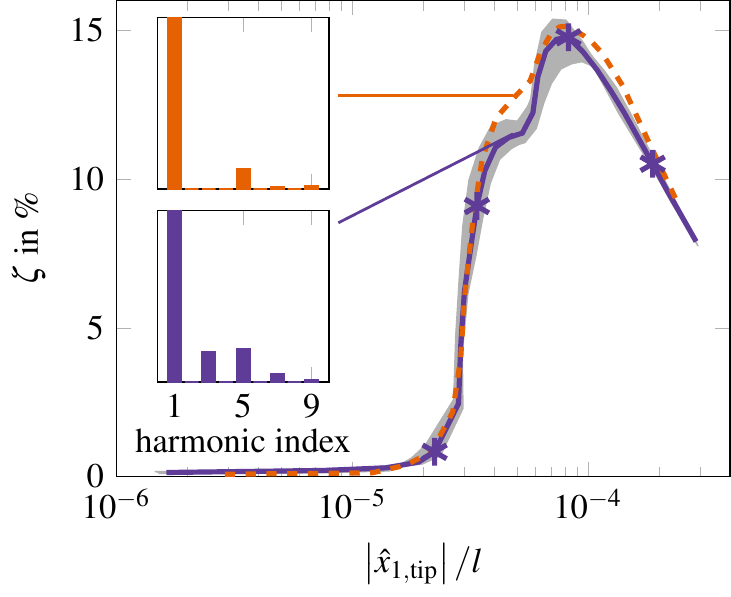}	
		\caption{}\label{fig:modal_properties_damp}
	\end{subfigure}
	\caption{(a) Amplitude-dependent modal frequency and (b) modal damping ratio for the excitation locations indicated in \fref{instrumentation}. The amplitude is the amplitude of the first Fourier coefficient of the beam's tip deflection, normalized with the beam's free length $l = 710 \text{ mm}$. The frequency is normalized with the modal frequency at low amplitudes (\cf \tref{linear_test}). The plotted curves are averages over six measurements and the gray-shaded area indicates the spread of measurements with excitation at the tip. The asterisks indicate the modal properties associated with the phase-resonant response at the four excitation levels of \fref{synthesis}. The harmonic content of the measured response is shown for one amplitude level.}
	\label{fig:modal_properties_freqdamp}
\end{figure}

To assess repeatability, one backbone was measured with increasing excitation level and, following immediately, with decreasing excitation level.
This procedure was conducted three times, leading to a total of six measurements, for each excitation location. Several tests were performed before the measurements presented here. This run-in time caused some wear in the contact area, but improved the repeatability of the subsequent measurements.
The air pressure was kept constant during all measurements of the same excitation location, \ie re-assembly repeatability was not assessed directly. The variation of the extracted frequency is below 2~\% or $\pm$ 1.5 Hz, and the variation of the tip acceleration is below 5~\%. This variation is independent of whether the excitation level is increased or decreased. In \fref{modal_properties_freqdamp}, the average of the six measurements is shown. Additionally, the spread of the measurements for excitation at the tip is indicated with the gray-shaded area.
The largest variation in the extracted frequency and in the damping ($\pm$ 0.6~\%) occurs for the amplitude range with maximal damping.

The averaged modal properties for excitation at two-third of the beam's length lie mostly in the repeatability spread of the tip excitation location. Therefore, we conclude that the extracted modal properties are independent of the chosen excitation location (\ie the differences in the shaker-structure-interaction caused by the different excitation locations).
This holds except for a small plateau in the damping curve obtained for excitation at the tip at about $5\cdot 10^{-5}$ normalized deflection, which is not observed for the other excitation location. Note that the plateau is repeatable for all backbones of this excitation location. A similar plateau was reported in a numerical study \cite{Krack2013}.
One potential cause for the plateau is a change in the higher-harmonic content of the motion. In \fref{modal_properties_damp}, the harmonic content of the measured response is shown for the two excitation locations at the amplitude range of the plateau. For excitation at the tip, the third and fifth harmonic are significant. For excitation at two-third of the beam's length, the third harmonic is not present in the response. 
Recall at this point that the ratio between the frequencies of the first two bending modes for low amplitudes is about 5.2. Due to the nonlinearity, the frequency of the second bending mode is also expected to depend on the vibration level. It is therefore likely that the two modes are in a 1:5 ratio for some vibration level. The plateau seen in the damping curve and the large contribution of the third and fifth harmonic could indicate an internal resonance. In the neighborhood of an internal resonance, the response is sensitive to the excitation.
Yet, it is unclear at this point if this difference in the harmonic content is in fact due to an internal resonance. The difference could also be caused by re-assembly variability, \ie a small change in the interface, since the air pillow was emptied and refilled between the measurements with excitation at the tip and excitation at two-third of the beam's length.

To analyze the nonlinear deflection shape, the amplitude-dependent first Fourier coefficient of the deflection is projected onto the linear modes,
\e{\bm \Gamma^{\text{(full-stick)}} = \shpmodlinmatrix^{+} \shpmodnorm_1.}{modal_projection}
This projection corresponds to the modal amplitude of the nonlinear deflection in the basis spanned by the linear modes.
Here, the first three linear bending modes of \tref{linear_test} are considered. The linear mode shapes are visualized in \fref{modes}. The absolute value of the components of $\bm \Gamma^{\text{(full-stick)}}$ is shown in \fref{modal_properties_shape_fixed}, normalized such that the sum of its entries is 1 for all amplitudes. Thus, a value of 1 indicates that the deflection shape is identical to the corresponding full-stick mode shape.
As expected, a clear change in deflection shape is observed as the similarity to the first (full-stick) mode decreases for higher amplitudes and the influence of the second and third full-stick mode increases for both excitation locations. The detected change in deflection shape is independent of the excitation location.
The computation of $\bm \Gamma$ is repeated with the bending modes identified without contact at the joint (see \fref{modal_properties_shape_free}). As expected, the nonlinear deflection shape for high amplitudes resembles the first bending mode without contact, indicated by an increasing value of $\abs{\Gamma_1^{\text{(free)}}}$ for higher amplitudes. In the range between $2\cdot 10^{-5}$ and $1\cdot 10^{-4}$ normalized deflection, there is a transition between the mode shapes of the two linear extreme cases, where the nonlinear deflection shape neither resembles closely the first full-stick nor the first bending mode without contact.
A similar analysis was conducted for the third and fifth Fourier coefficient. The second linear mode contributes significantly to these higher harmonics. This contribution of two linear modes and, at the same time, strong damping is identical to the limitations of the usefulness of \EPMC.

\begin{figure}
	\centering
	\includegraphics{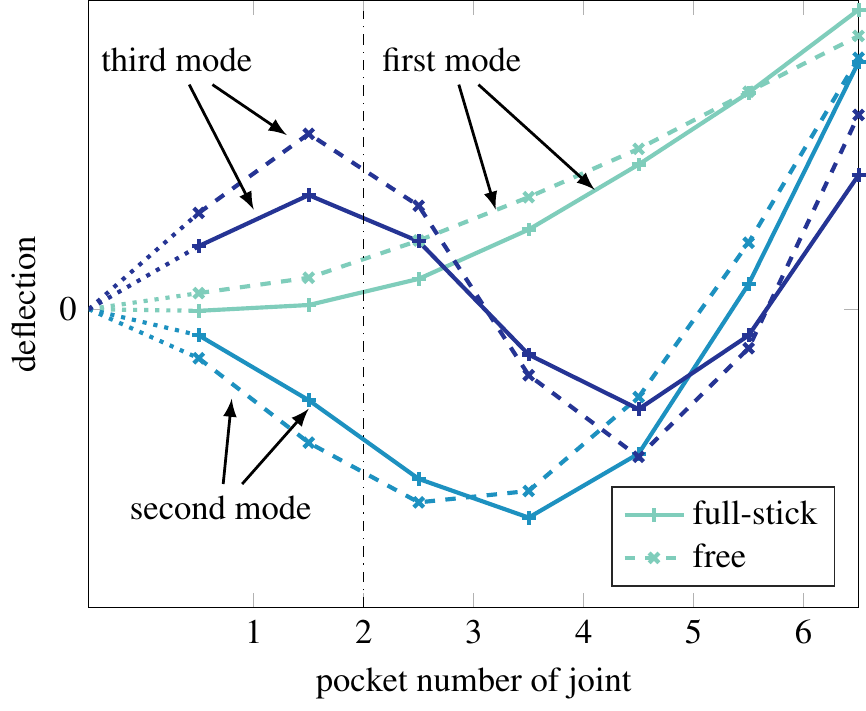}
	\caption{First three linear bending modes in horizontal direction for full-stick contact (dashed lines) and without contact (free) at the joint (solid lines). The dotted lines indicate the estimated deflection between clamping and first sensor.}
	\label{fig:modes}
\end{figure} 

\begin{figure}
	\centering
	\begin{subfigure}{0.49\textwidth}
		\centering
		\includegraphics{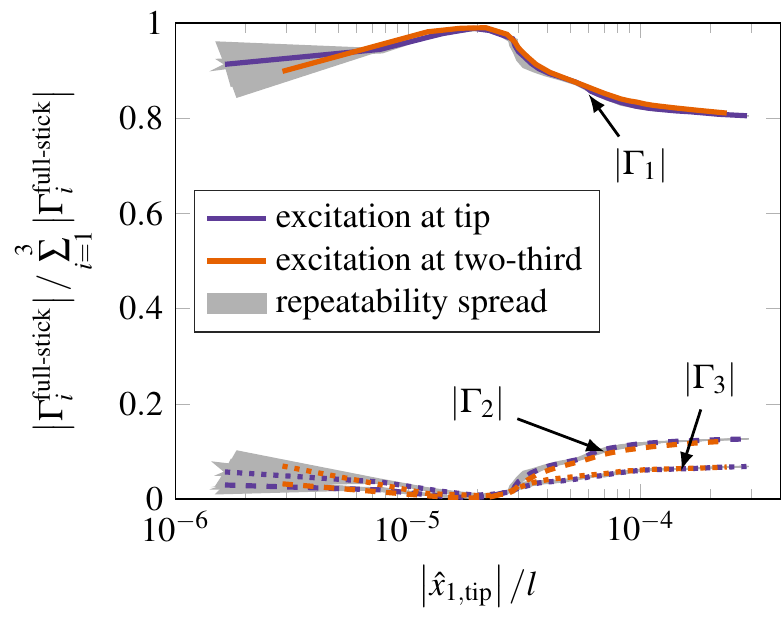}
		\caption{}\label{fig:modal_properties_shape_fixed}
	\end{subfigure}
	\begin{subfigure}{0.49\textwidth}
		\centering
		\includegraphics{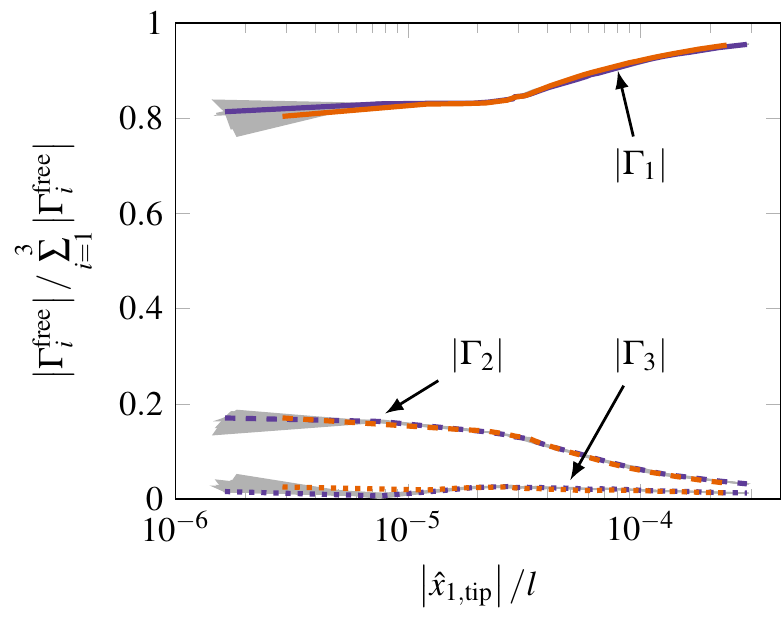}
		\caption{}\label{fig:modal_properties_shape_free}	
	\end{subfigure}
	\caption{Normalized absolute value of the modal amplitudes of the first (solid line), second (dashed line), and third (dotted line) linear bending modes (a) with contact at the joint and (b) without contact. The index indicates the corresponding linear mode index. The gray-shaded area indicates the repeatability spread of the deflection shape obtained with excitation at the beam's tip.}\label{fig:modal_amp}
\end{figure}

Another way to visualize the nonlinear deflection shape is to plot the motion in the configuration space (the space spanned by generalized coordinates). In \fref{loops}a-c, the deflection $x_{\text{tip}}$ of the tip sensor is plotted against the deflection $x_{\text{pocket 2-3}}$ of the sensor between pockets two and three (close to the joint) for seven representative points on the backbone. The deflection is obtained by trapezoid integration of the measured acceleration, assuming zero-mean acceleration and velocity.
The loops in the configuration space indicate non-synchronous motion, \ie a non-trivial phase lag between the deflection of the two sensors. Moreover, the "waviness" of the loops indicates the presence of higher-harmonic content. For some levels, crossing loops are observed. Finally, the shape of the configuration space loops differs considerably for different vibration levels indicating substantial local changes in the deflection shape. These changes can also be observed when visualizing the phase space. In \fref{phase_space}, the normalized velocity is plotted against the normalized deflection of the sensor between pocket two and three. For the lowest excitation level, the motion resembles an ellipse. For higher levels, the higher harmonics are apparently present in the trajectories. They distort the elliptic shape.

\begin{figure}
	\centering
	\begin{subfigure}{0.45\textwidth}
		\includegraphics{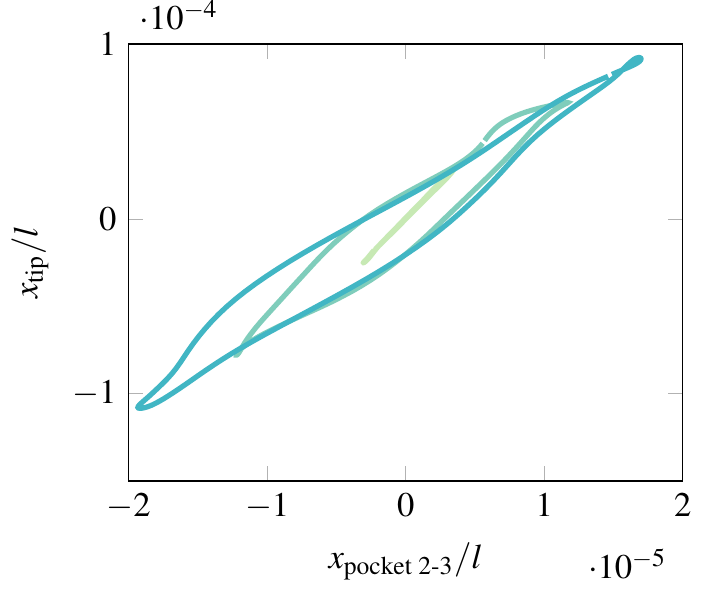}\caption{}
	\end{subfigure}
	\begin{subfigure}{0.45\textwidth}
		\includegraphics{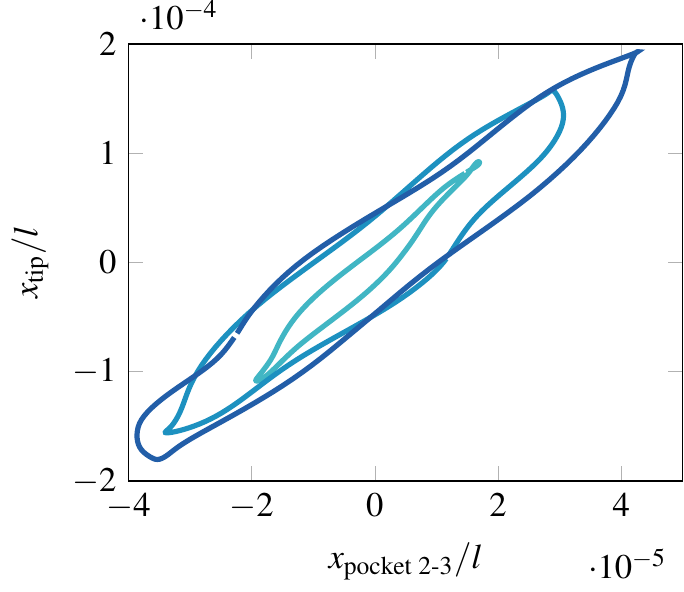}\caption{}
	\end{subfigure}
	\begin{subfigure}{0.45\textwidth}
		\includegraphics{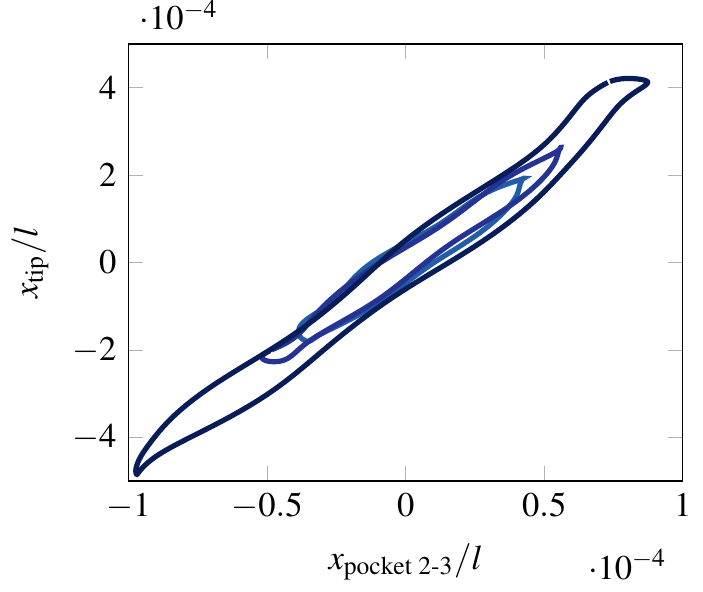}\caption{}
	\end{subfigure}
	\begin{subfigure}{0.45\textwidth}
		\includegraphics{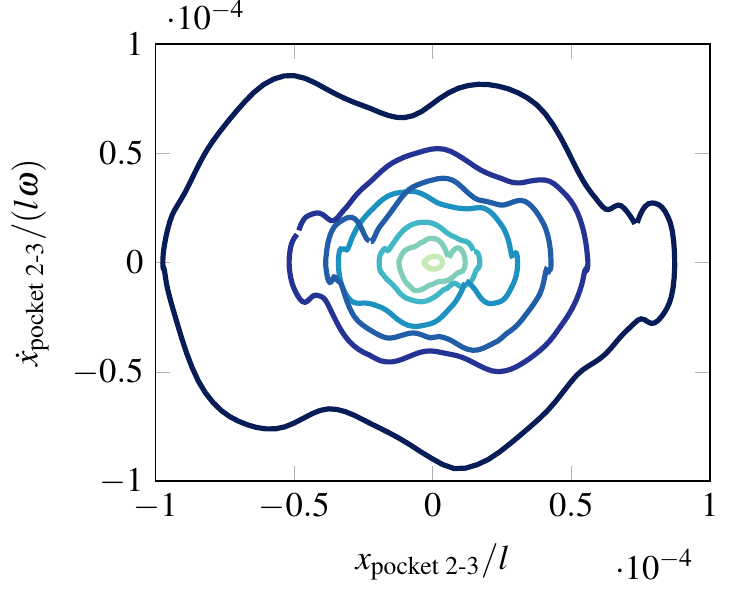}
		\caption{}\label{fig:phase_space}
	\end{subfigure}
	\caption{(a)-(c) Configuration space illustration of the deflection shape. The normalized deflection at the beam's tip is plotted against the normalized deflection between pocket two and three (close to the joint) for seven different excitation amplitudes. (d) Phase space projection of the same cycles of vibration. The normalized velocity between pocket two and three is plotted against the normalized deflection at the same location.}\label{fig:loops}
\end{figure}

Recall at this point that the simplified forcing described in \sref{forcing} only ensures phase resonance with respect to the fundamental harmonic. It is well known that the shaker-structure interaction can cause higher harmonics in the excitation force (which are not controlled in the present implementation of the nonlinear modal testing method).
In the presented experiments, the measured excitation force is dominated by the fundamental harmonic with a total harmonic distortion (with respect to the root mean square value \cite{Shmilovitz2005}) mainly between 2 \% - 4 \% and maximum 7 \% at some excitation levels.
Yet, the higher harmonics of the vibration response are significant (total harmonic distortion mainly between 10 \% - 30 \% for the measured acceleration signal). They are neither synchronous with the fundamental harmonic nor in a 90$^\circ$ phase lag with the respective higher harmonics of the excitation force. These uncontrolled harmonics affect the excitation force (through the shaker-structure interaction) and thus deteriorate the excitation of the mode. To improve the mode isolation, multi-harmonic controlled forcing could be utilized.

Analyzing the wear traces at the inserts after testing, one can observe that the normal load is not distributed equally over the four lines. On the bottom two bosses (\fref{wearb}), there is no contact in the mid of the insert. At the top two bosses (\fref{weart}), there is full contact on one boss and only contact on half of the length of the other boss. The unequal normal preload distribution is probably caused by machining tolerances. Using a different insert, the contact situation would most likely be different.

\begin{figure}
	\centering
	\begin{subfigure}{0.45\textwidth}
		\includegraphics[width=.9\textwidth]{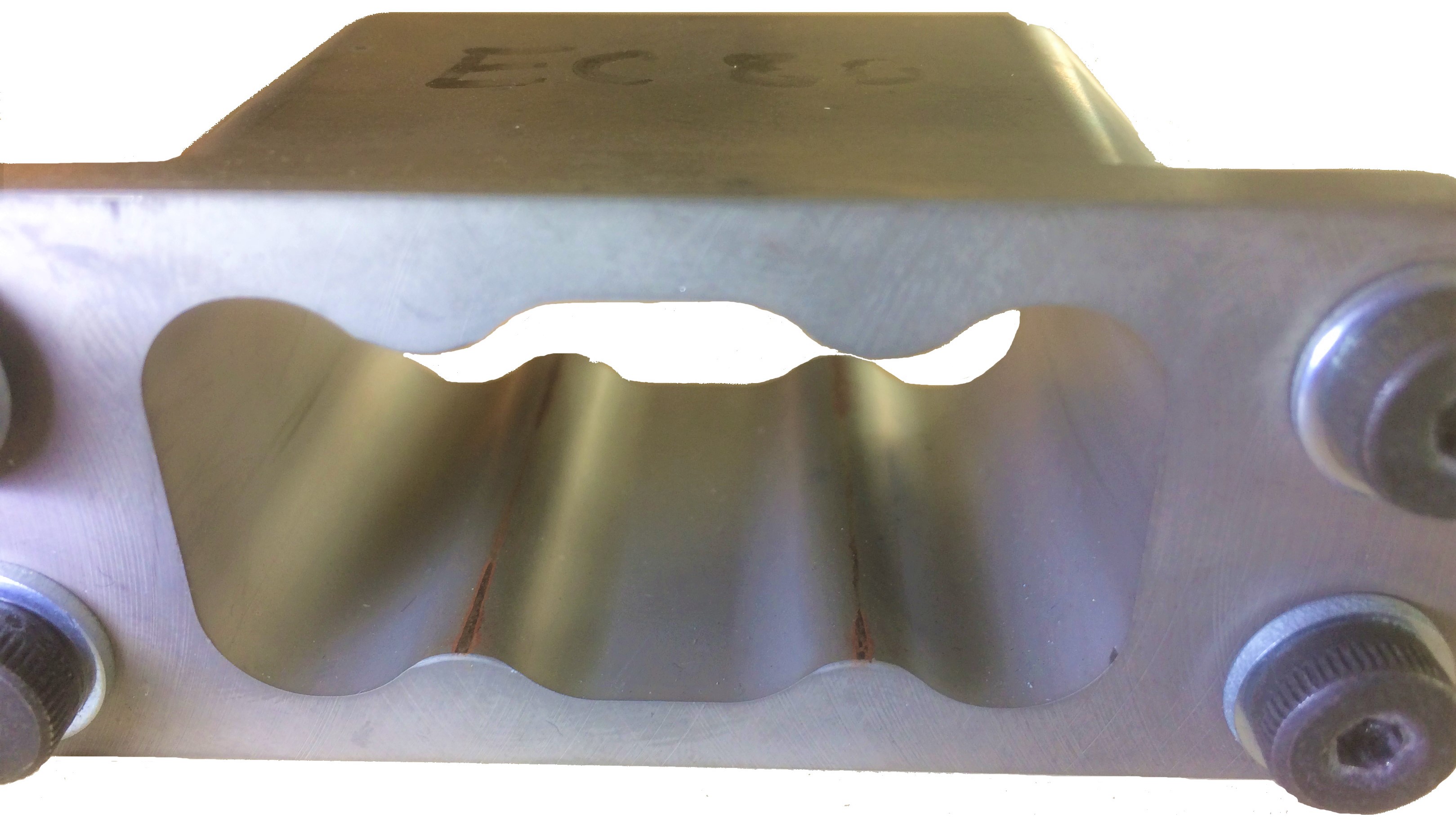}	
		\caption{}\label{fig:wearb}
	\end{subfigure}
	\begin{subfigure}{0.45\textwidth}
		\includegraphics[width=.9\textwidth]{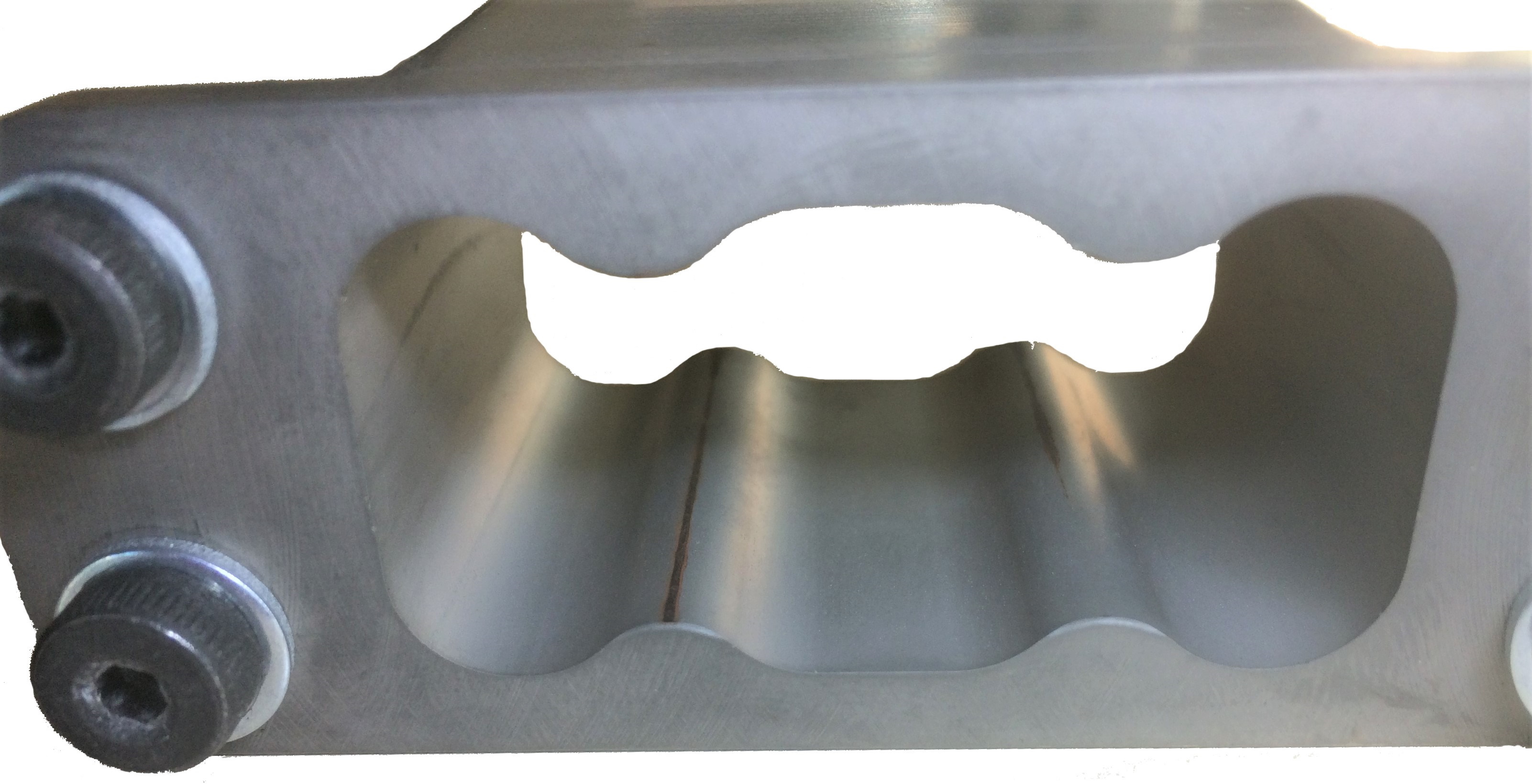}
		\caption{}\label{fig:weart}	
	\end{subfigure}
	\caption{Wear of the (a) bottom and (b) top bosses of the insert after testing.}
\end{figure}

\subsection{Validation of the Amplitude-dependent Modal Properties}

According to the \EPMC, the amplitude-dependent modal properties can replicate dynamics close to an isolated resonance. To assess the quality of the extracted modal properties, steady-state frequency responses are computed around the resonance and compared with measured responses.
As reference, controlled stepped sine measurements with excitation at the tip were conducted. To this end, the PLL controller was utilized to control stepped phase lags between 20$^\circ$ and 145$^\circ$.
The fundamental harmonic force amplitude was controlled to a constant level with an additional outer control loop using a PI controller with gains $K_i = 0.5 \; \mathrm{V}\: \mathrm{N}^{-1} \: \mathrm{s}^{-1}$ and $K_p = 0.5  \; \mathrm{V}\: \mathrm{N}^{-1}$. Four fundamental harmonic amplitudes of the excitation force were tested, namely 0.40 N, 5.03 N, 14.55 N and 20.35 N.
The PLL control parameters were $K_p = 10 \; \mathrm{s}^{-1}$ and $K_i = 5\pi \; \mathrm{s}^{-2}$ for forcing levels 5.03 N and 14.55 N. For forcing levels 0.40 N and 20.35 N, they were $K_p = 20  \; \mathrm{s}^{-1}, K_i = \pi \; \mathrm{s}^{-2}$ and $K_p = 15 \; \mathrm{s}^{-1}, K_i = 5\pi  \; \mathrm{s}^{-2}$, respectively. All control parameters were chosen heuristically.

For each excitation level, three frequency responses were measured directly one after another, which were well repeatable.
The average of the measurements are shown with black dots in \fref{synthesis}.
Note that two of the six backbone measurements were conducted before and the other four backbones after the frequency response measurements. Thus, potential changes in the dynamics due to changes in the contact interface during the reference measurements are included in the repeatability assessment of the backbones.

The frequency response of a harmonically forced nonlinear modal oscillator is governed by
\ea{[- \Omega^2+2 \ii \Omega \ommod(\modamp) \Dmod(\modamp) + \ommod(\modamp)^2] \modamp \ee^{\ii\phasemod} = \shpmodnorm\herm_1 (\modamp) \ForceVec_{1,\rm exc}.
}{nmsdof}
$\ForceVec_{1,\rm exc}$ and $\Omega$ are the fundamental Fourier coefficient and angular frequency of the excitation force.
This equation is solved explicitly for $\Omega^2$ at the amplitude levels covered by the backbone measurements \cite{Schwarz2019}.
To increase the resolution of the prediction, the modal properties have been interpolated with piecewise cubic Hermite polynomials, which were found useful in the past \cite{Scheel2020}.
The forced responses for four excitation levels are shown in \fref{synthesis}, taking into account harmonic forcing with the same level and location as the reference measurements. The modal properties at the corresponding phase-resonant points of these force levels are marked with asterisks in \fref{modal_properties_freqdamp}. The gray-shaded areas indicate the predictions' spread using the individual six backbone measurements with excitation at the tip.

Both nonlinear-modal oscillators identified with data of the two excitation locations predict the frequency responses well for all four excitation levels. Therefore, the accuracy of the extracted modal properties is demonstrated.
For the two lowest and the highest excitation level, the reference measurements are within the repeatability spreads. Thus, these frequency responses are predicted accurately, considering the natural spread of the measurements.
For the second highest level, the predicted response captures the frequency response's shape well, though the predicted response is slightly shifted to lower frequencies. For frequency ratios higher than 0.8, the reference measurement shows a local maximum, which the model cannot predict. This local maximum was observed in all reference measurements of this level.
In the frequency range around this local maximum, the influence of the higher harmonics in the response is large compared to other frequencies of the same level. 
Recall that the system is strongly damped for the vibration level of the second highest excitation level, and, at the same time, two linear modes contribute to the motion. This indicates a modal interaction, where the assumption of a single nonlinear mode is no longer valid. A double peak in the frequency response can be caused by modal interactions, as was observed in numerical studies (see \eg \cite{Krack2013}).

\begin{figure}
	\centering
	\includegraphics{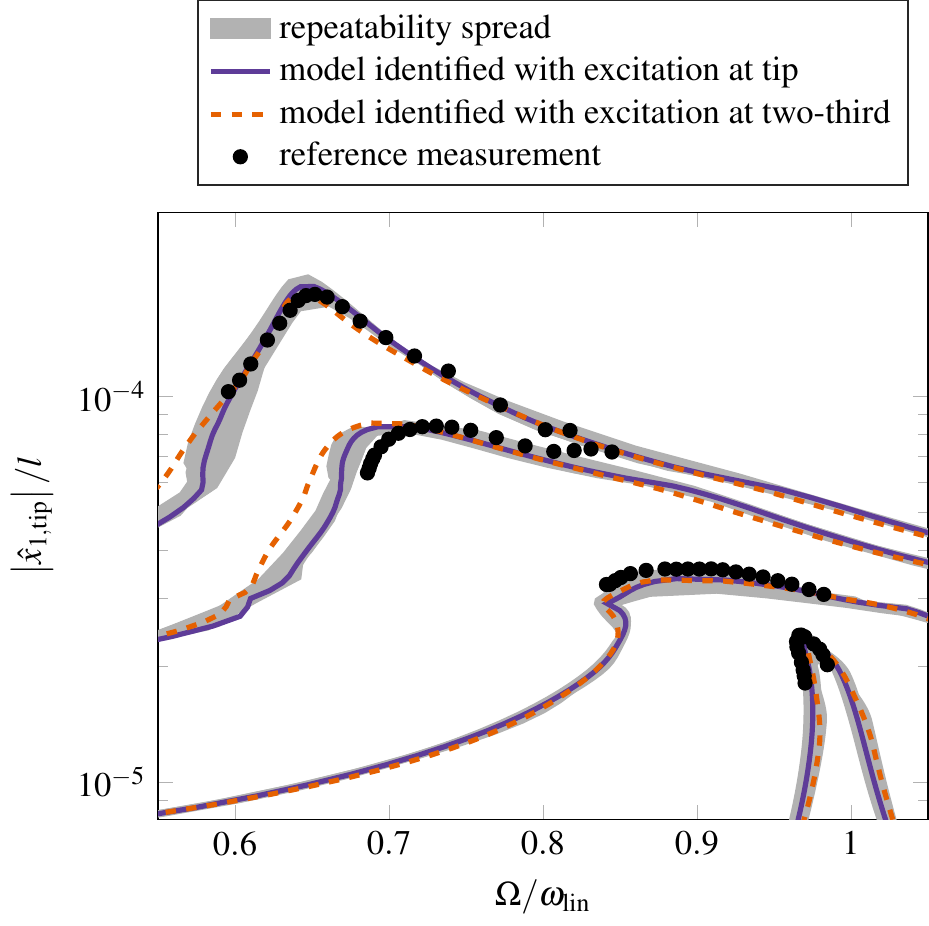}
	\caption{Predicted steady-state frequency responses for four excitation levels with harmonic forcing at the beam's tip. The excitation levels are 0.40 N, 5.03 N, 14.55 N and 20.35 N. The gray-shaded areas indicate the spread of the individual six backbone measurements' predictions with excitation at the tip. The frequency responses are shown for the normalized deflection of the beam's tip and the frequency axis is normalized with the modal frequency at low amplitudes (\cf \tref{linear_test}).}
	\label{fig:synthesis}
\end{figure}

\section{Conclusions}\label{sec:conclusion}

In this work, we present a new test rig called RubBeR. It is a beam strongly influenced by friction damping ranging from full stick to mainly sliding.
The modular design of the test rig allows for varying the location of the joint as well as the contact partners of the friction interface. 
The configuration studied in this paper exhibits a decrease in frequency by 36 \% and an increase in the modal damping ratio to about 15 \% for the first horizontal bending mode. The spatial distribution of the deflection shape changes with amplitude, resembling the respective linear bending mode for the limit cases at low and high amplitudes. Moreover, the contribution of higher harmonics to the deflection shape changes with vibration amplitude. The modal properties are well repeatable for a system subjected to dry friction. These properties make RubBeR a suitable and challenging test structure for the assessment of nonlinear system identification methods.

Two sets of modal properties are obtained from measurements with single-point forcing using shaker-stinger excitation at two different locations. The respective modal properties are in very good agreement, considering the spread of repeated measurements.
The analysis of the measured backbone signals shows that significant, uncontrolled higher harmonics are present in the response.
The accuracy of the modal properties is assessed by computing frequency responses to harmonic forcing using a single nonlinear-modal oscillator with the modal properties identified with the modal analysis. These frequency responses agree well with measured reference curves of different excitation levels. Therefore, we conclude that the applied nonlinear modal testing method with phase control using a simplified excitation mechanism with single-point, single-harmonic forcing is sufficient to extract meaningful modal properties for this strongly nonlinear test specimen.
In this spirit, we motivate future work  on multi-point and multi-harmonic controlled excitation, by analyzing to what extent the mode isolation quality could be improved.


\section*{Acknowledgment}
 
This work was funded by the Deutsche Forschungsgemeinschaft (DFG, German Research Foundation) [Project 402813361].
The authors would like to thank Stefan Schwarz, Daniel Fochler and Markus Klein for their valuable feedback and ideas concerning the design of RubBeR.

\bibliography{Rubber_paper_ref}

\end{document}